\documentclass{optica-article}

\journal{opticajournal} 

\articletype{Research Article}

\usepackage{lineno}

\usepackage{graphicx}
\usepackage{dcolumn}
\usepackage{bm}
\usepackage{amsmath,physics,color,enumerate,tabularx,makecell,mathrsfs,float,bbm,stackrel,mathtools,multirow,bm}
\usepackage{xcolor}
\usepackage{comment}
\usepackage{caption}
\usepackage{subfig}
\usepackage[normalem]{ulem}
\begin{document}

\title{Quantum steering from phase measurements with limited resources}

\author{Gabriele Bizzarri,\authormark{1} Ilaria Gianani,\authormark{1}, Mylenne Manrique,\authormark{1}, Vincenzo Berardi,\authormark{2}, Fabio Bruni,\authormark{1}, Giovanni Capellini,\authormark{1}, and Marco Barbieri\authormark{1,3,*}}

\address{\authormark{1}Dipartimento di Scienze, Universit\`a degli Studi Roma Tre, Via della Vasca Navale 84, 00146 Rome, Italy\\
\authormark{2}Dipartimento Interateneo di Fisica “M. Merlin”, Politecnico di Bari, Via Orabona 4, 70126 Bari, Italy
\\
\authormark{3}Istituto Nazionale di Ottica - CNR, Largo E. Fermi 6, 50126 Florence, Italy
}

\email{\authormark{*}marco.barbieri@uniroma3.it} 


\begin{abstract*} 
Quantum steering captures the ability of one party, Alice, to control through quantum correlations the state at a distant location, Bob, with superior ability than allowed by a local hidden state model. Verifying the presence of quantum steering has implications for the certification of quantum channels, and its connection to the metrological power of the quantum state has been recently proved. This link is established by means of the violation of a Cram\'er-Rao bound holding for non-steerable states: its direct assessment would then require operation in the asymptotic regime of a large number of repetitions. Here, we extend previous work to account explicitly for the use of a limited number of resources, and put this modified approach to test in a quantum optics experiment. The imperfections in the apparatus demand an adaptation of the original test in the  multiparameter setting. Our results provide guidelines to apply such a metrological approach to the validation of quantum channels.
\end{abstract*}

\section{Introduction}
Quantum correlations are responsible for most peculiar quantum effects, yielding to the perspective of communications with unprecedented security~\cite{RevModPhys.94.025008} and of sensing with unprecedented precision~\cite{PhysRevLett.96.010401}. Different types of quantum correlations exist, each capturing a different aspect and associated to a communication scenario with varying degree of trust among the nodes~\cite{PhysRevA.85.010301}. The presence of mere quantum entanglement, defined as the nonseparability of the shared quantum state, is sufficient to connect mutually trusting nodes. Quantum nonlocality, instead, is necessary in order to lift assumptions on the functioning of the sources, as well as of the detection apparatus\cite{PhysRevLett.113.140501}. 

At the intermediate level there lays quantum steering ~\cite{PhysRevLett.98.140402,RevModPhys.92.015001,PRXQuantum.3.030102}. This kind of correlation is associated to the ability of one party, Alice, to control the state at the receiving end, Bob, beyond what is possible by using only local quantum states taken with a given probability distribution. This concept has been the subject of numerous studies, opening the path to understanding its implications for quantum technologies~\cite{Saunders:2010xw,PhysRevX.2.031003,Handchen:2012jl,Smith:2012zt,Wittmann_2012, Cavalcanti:2015qy,Kocsis:2015ee,PhysRevLett.116.160404,Westone1701230,PhysRevLett.121.100401}. Yadin, Fadel and Gessner (YFG) have recently established a formal connection between the presence of quantum steering in a state and its metrological capabilities, as captured by the Fisher information~\cite{YBG}. This is part of a wider effort to establish synergies between quantum communications and quantum sensing and metrology~\cite{PhysRevA.99.022314,PhysRevA.105.L010401}.  

An implementation of the YFG test has been carried out by a direct assessment of the Fisher Information~\cite{PhysRevA.105.022421}. Such an approach benefits from an immediate link to the YFG treatment, and the characterization of the measuring device naturally encompasses sources of errors. This route demands a large number of copies of the steerable state for a satisfactory retrieval of the Fisher information from the analysis of the measurement. This latter requirement constitutes a limiting factor for its adoption in communication networks, thus suggesting to turn one's attention to the precision in parameter estimation. The uncertainty on a given parameter can be assessed with fewer resources, and one can then make use of the celebrated Cram\'er-Rao bound (CRB) to connect the retrieved uncertainty to the Fisher Information. However, imperfections affect the estimation, and, since these pertain to the quantum state, they have to be included as extra nuisance parameters. Hence, the multiparameter estimation approach is needed. In addition, the CRB provides a reliable assessment of the precision only when a large ensemble is used for the estimation. Consequently, far from this regime, finite data-size effects should be taken into account~\cite{Gianani20}.  These effects have also attracted attention in the context of quantum key distribution~\cite{PhysRevA.81.062343,Coles16,Bunandar2020}. 

This work reports on the investigation of these concepts in a quantum optics experiment. We witness the presence of quantum steering by comparing the precision of phase estimation at Bob's location with a limiting bound obtained by extending the treatment of YFG to the multiparameter scenario and to the finite-ensemble regime. Our results indicate the viability of this approach, and pave the way for its inclusion in the toolbox of nonclassicality tests for shared resources, especially when these are intended for distributed quantum sensing.   

\section{Results}

Two parties, Alice and Bob, share a quantum state $\rho^{AB}$. The presence of quantum correlations in $\rho^{AB}$ results in the fact that a measurement of the observable $K$ on Alice's side steers the state in Bob's location. This makes her able to guess the outcome of a measurement of Bob's observable $G$. Bob accepts the validity of quantum mechanics for the state of his own partition, still he does not believe Alice can steer it better than a local hidden state (LHS) theory allows. This predicates that Bob's measurement outcomes are obtained by applying Born's rule to a LHS $\sigma^B_k(\lambda)$, associated with Alice's value $k$ for $K$. The presence of a variable $\lambda$ is assumed, and it governs both the probability distribution of Bob's local states, and the probability $p(k\vert K)$ of Alice  observing the outcome $k$, given the choice $K$ for her observable. These two aspects find a unified description with the introduction of the {\it assemblage} $\mathcal{A}(k,K)=p(k\vert K)\rho^B_{k\vert K}$, where $\rho^B_{k|K}$ is Bob's conditional state. Here $k$ is treated as a classical variable, and no assumptions are made about its connection to a quantum measurement at Alice's side.  

If a LHS theory is suited for the description of $\rho^{AB}$, all assemblages could be written as $\mathcal{A}(k,K)=\int_\lambda p(k\vert K,\lambda) \sigma^B_k(\lambda)d\mu(\lambda)$. The adequacy of this expression is usually tested by constructing a steering parameter $S=\sum_i \langle k_i G_i\rangle$, containing correlations between Alice's classical variables and Bob's quantum observables for distinct settings $i=1,...,n$.  The LHS model provides a limit $C$ such that a violation of the condition $S\leq C$ testifies the presence of quantum steering~\cite{Saunders:2010xw}, in analogy with Bell's test for nonlocality. A different take on witnessing steering considers the conditional variance for different choices of $G$; this is defined as $\Delta^2G_{\rm cond} = \sum_{k,g} p(k\vert K)\Delta^2G[\rho^B_{k|K}] $. For any two observables $G$ and $G'$, a violation of the condition $\Delta^2G_{\rm cond} \,\Delta^2G'_{\rm cond}\geq \langle [G,G'] \rangle/4$ leads to excluding local hidden state models. This result is known as Reid's criterion~\cite{PhysRevA.40.913,PhysRevA.80.032112}. 

\begin{figure}
\includegraphics[width=\columnwidth]{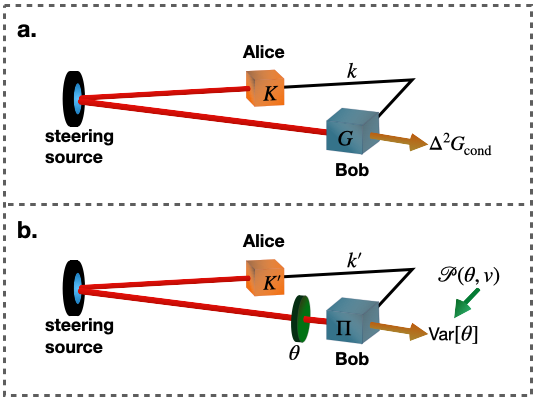}
\caption{Metrological test of quantum steering. a. Alice communicates to Bob a classical message $k$ she derived from a measurement of her observable $K$. Bob uses such events to infer the conditional variance of his observable $G$. b. Bob inserts an object whose characteristic parameter $\theta$ has to be estimated. The uncertainty $\text{Var}[\theta]$ is obtained from the measurement outcomes of a POVM $\Pi$ and from the prior probability distribution $\mathcal{P}(\theta,v)$ for the target parameter $\theta$, as well as for the nuisance parameter $v$.}
\label{fig:concept}
\end{figure}

The work of YFG has established the connection between Reid's criterion and an estimation protocol, illustrated in Fig.~\ref{fig:concept}. Bob uses his substate to estimate a parameter $\theta$ appearing in a unitary transformation $U_\theta =e^{-i\theta G }$. He then applies $U_\theta$  and performs a measurement on $N$ copies, thus recovering the parameter within an uncertainty quantified by the variance of its estimator $\text{Var}[\theta]$. There exists a quantity called Fisher information $f(\theta)$ that is associated with the probability of the measurement outcomes, thus, in our case, to Alice\sout{s}'s and Bob's choices of their observables. The relation $N\text{Var}[\theta]f(\theta)\geq 1$, known as CRB~\cite{GLMreviewScience}, is established as a fundamental theorem in metrology (see Supplemental Material).  

The central result of YFG is that the Fisher information $f(\theta)$ for measurements on states admitting a  LHS model must satisfy $f(\theta)\leq \Delta^2 G_{\rm min}=\min_K \sum_k p(k|K)\Delta^2G[\rho^B_{k|K}]$. Noticing that by definition $\Delta^2 G_{\rm cond}\geq \Delta^2 G_{\rm min}$, we can find a chain of inequalities leading to 
\begin{equation}
\label{eq:crb}
    \text{Var}[\theta] \Delta^2 G_{\rm cond} \geq \frac{1}{4N  }.
\end{equation}

This scenario, however, assumes that Bob has all the means to get a value for the estimator of $\theta$ from his data, thus implying there are no additional parameters to be known. This is hardly the case in the real scenario: there will be the need of introducing further quantities for an effective description of the outcome statistics. All of these are included in a vector of parameters~\cite{Suzuki_2020,PhysRevA.105.012411}. In the simplest case, the target parameter $\theta$ is complemented by a single nuisance parameter $v$, generally associated with a dissipative process. The appropriate tool to assess uncertainties is now the covariance matrix of $\theta$ and $v$, which the multiparameter CRB limits by means of a Fisher information matrix $F$. 

In addition, the strictness of the CRB is also influenced by the number of copies $N$, as this should be large in order to suppress significant deviations of $\text{Var}[\theta]$ from its expected value. When this condition is not met, it is not fair to assume all the information on the parameters is obtained uniquely from the measurements.
Therefore, a prior probability distribution $\mathcal{P}(\theta,v)$ must be introduced  explicitly in order to account for all of Bob's previous knowledge. Its availability is actually implicit in the local estimation framework for which the CRB is relevant~\cite{ParisReview}, but its role cam be overlooked in the large-ensemble limit. An extension of the CRB, due to Van Trees, accounts for this additional source of information in the final variance~\cite{ParisReview,Gill95}.

Thanks to Van Trees' treatment, these considerations on the multiparameter setting as well as on the limited-resource regime can be encompassed in a modified version of \eqref{eq:crb}, reading 
\begin{equation}
\label{eq:bizzarri}
        N\text{Var}[\theta] \mathcal{L} \geq 1,
\end{equation}
with
\begin{equation}
   \mathcal{L} = 4\Delta^2 G_{\rm cond}+\frac{1}{N} \int \frac{\left( \partial_\theta \mathcal{P}(\theta,v)\right)^2}{\mathcal{P}(\theta,v)} d\theta d v.
\end{equation}
A detailed proof is given in the Supplemental Material. All quantities in \eqref{eq:bizzarri} can be assessed in the experiment. We remark how, differently than in the standard setting, the variance is not compared to a theoretical limit, but rather to a quantity obtained by distinct, independent measurements. 

\begin{figure}[b]
\includegraphics[width=\columnwidth]{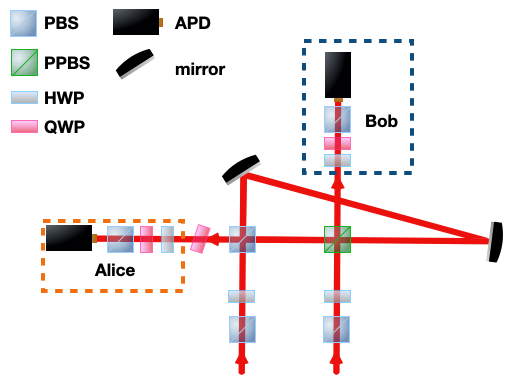}
\caption{Experimental setup. States exhibiting quantum steering are produced by means of a control-Z gate applied to two polarization qubits. The two photons are prepared in a suitable polarization state in order to produce a state close to the maximal entanglement. The gate is embedded within a Sagnac arrangement with the two paths showing different polarizations. An extra QWP compensating the resulting phase is therefore needed. The analysis is performed by a standard tomographic setup. The setup is in free space, but photons are delivered to it by means of single-mode fibers.
Likewise, the outputs are fiber-coupled to avalanche photodiodes (APDs) detectors.}
\label{fig:setup}
\end{figure}

We have tested these concepts with two-qubit states as shown in the setup of Fig.~\ref{fig:setup}. 
A CW parametric downconversion source produces degenerate photon pairs at $\lambda =810$ nm with bandwidth $7.5$ nm imposed by interference filters. Both photons are initially prepared in the state $\vert \tilde{D}\rangle =\cos\left({\pi/3}\right)\vert H \rangle + \sin\left({\pi/3}\right) \vert V \rangle$ - where the ket $\vert H \rangle$ $\left(\vert V \rangle\right)$ denotes a photon with horizontal (vertical) polarization. The two photons are then sent on a partially polarizing beam splitter (PPBS) with transmittivities $T_H=1$, $T_V=1/3$, embedded in a Sagnac interferometer to reduce the effects of the PPBS imperfections~\cite{Cimini20}. Polarization-selective nonclassical interference results in the production of an entangled state in the form $\ket{\psi}_{AB}=2^{-1/2}\left(\ket{HD}_{AB}+\ket{VA}_{AB}\right)$, with $\ket{D}=2^{-1/2}\left(\ket{H}+\ket{V}\right)$, $\ket{A}=2^{-1/2}\left(\ket{H}-\ket{V}\right)$ when post-selecting on coincidence events. The polarization-dependent loss at the PPBS is responsible for the balanced populations~\cite{PhysRevLett.95.210506}. Alice and Bob can then perform measurements of the Pauli Operators $X$,$Y$ and $Z$ of their qubits by means of a half wave plate (HWP), a quarter wave plate (QWP) and a polarizer. As customary, the observable $Z$ is associated to the $H/V$ basis, $X$ to the diagonal $D/A$  basis, and finally $Y$ to the circular basis. In our implementation, Bob's parameter is a rotation angle $\phi$ of its polarization by means of $U=e^{-i\phi\frac{Y}{2}}$, {\it i.e.} $G=Y/2$. The target phase $\phi=\pi/4$ can then be imparted simply by adjusting the angular position in Bob's measurement HWP. Imperfections affect the bipartite state, mostly the limited contrast in nonclassical interference. Bob can account for these imperfections by adding an effective visibility $v$ of his fringes acting as the additional nuisance parameter he has to estimate~\cite{Roccia:18}.  

The protocol then proceeds as follows: when Alice measures her observable $Z$, Bob performs the joint estimation of the phase $\phi$ and the visibility $v$ through a generalised measurement alternating between $X$ and $Z$ with equal weights~\cite{PhysRevA.81.012305}, allocating $N = N_Z$ resources to this task. When Alice measures $Y$, Bob also measures $Y$ in order to obtain the value of $4\Delta^2G_{\rm cond}=\Delta^2Y_{\rm cond}$ and assess a bound on this Fisher information, this time based on $N_Y$ repetitions.

\begin{figure}
\includegraphics[width=\columnwidth]{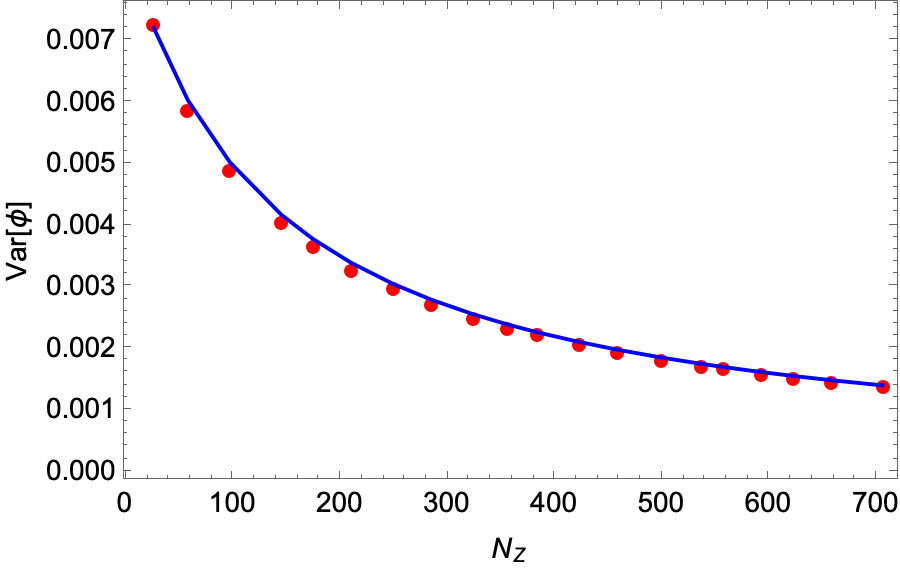} 
\caption{Variance for the estimation of $\phi$. For each value of $N_Z$, a Bayesian estimator is used to derive a distribution for the parameter $\phi$, from which the variance is calculated and reported here as the red points. The blue solid curve is the Van Trees limit.
} 
\label{fig:phase_N}
\end{figure}

The results for the phase estimation uncertainties $\text{Var}[\phi]$ are reported in Fig.~\ref{fig:phase_N} as a function of the number of events $N_Z$ used in the protocol. These follow the trend predicted by Van Trees's bound~\cite{Gill95}, shown as a solid line.  
This bound extends the conventional Cram\'er-Rao limit to include the informational content from the prior probability distribution $\mathcal{P}(\phi,v)$, with a weight decreasing as $N_Z$ increases. A slight deviation from the bound is observed which we may ascribe to effects not included in our model - foremost a slight unbalance in the populations with respect to the equal superposition in the state $\ket{\psi}$. These uncertainties are employed to verify a violation of \eqref{eq:bizzarri}, with the term $\mathcal{L}$ constructed by means of a direct evaluation of $\Delta^2Y_{\rm est}$ and from our choice of the prior distributions. 
The continuity and derivability requirements for the Van Trees bound to be applied demand a cautious choice of the prior distributions.
Furthermore, the prior distribution for $\phi$ should be sufficiently narrow to disambiguate the value within the periodicity. For the purposes of our analysis, it is important to stress that $\mathcal{L}$ is determined within the experimental error on $\Delta^2Y_{\rm cond}$.

The variable $\xi^2=N_Z\text{Var}[\phi]\mathcal{L}$ is a normalized variance, hence it is expected to follow the distribution for a normalized $\chi^2$ variable. We can thus formulate our steering criterion as a one-sided test with $N_z-1$ degrees of freedom, for which Eq. \eqref{eq:bizzarri} is the null hypothesis: critical values $\chi^2_{{\rm cr},p}$ can be then calculated corresponding to different confidence intervals. The standard procedure is then completed by verifying that measured value of $\xi^2$ does not satisfy $\xi^2\geq \chi^2_{{\rm cr},p}$. 

\begin{figure*}
\includegraphics[width=0.5\columnwidth]{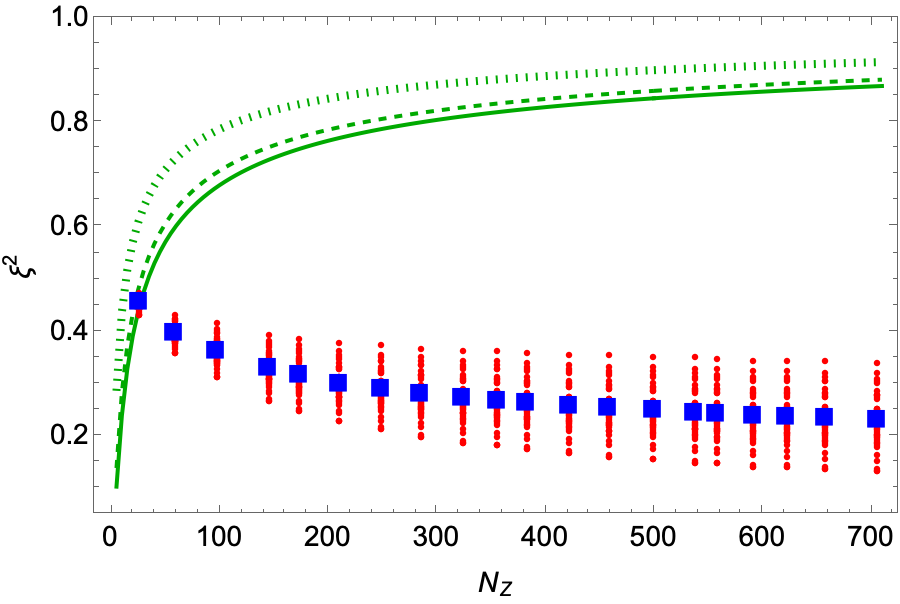}
\includegraphics[width=0.5\columnwidth]{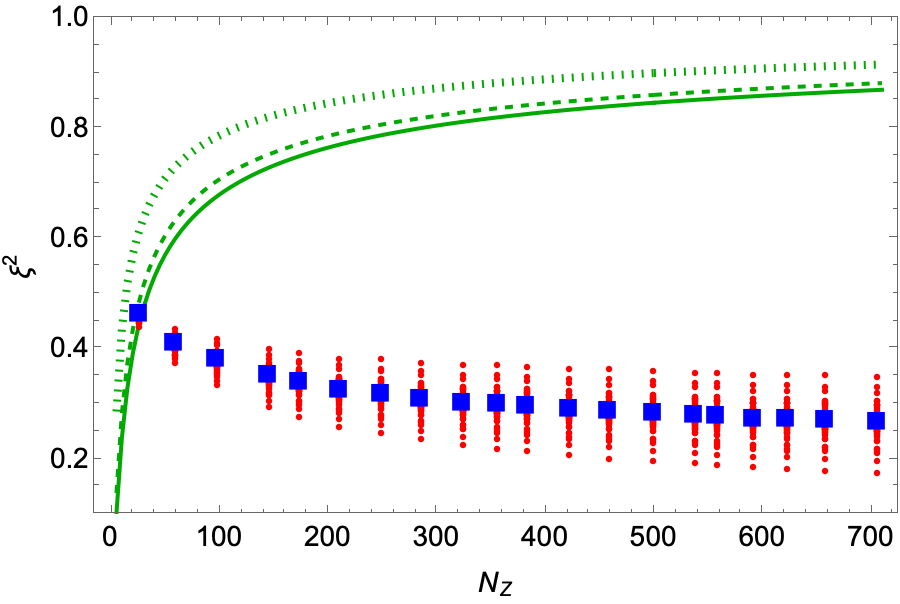}
\caption{Validation of the violation of the YFG limit \eqref{eq:bizzarri}. In both panels, the blue points refer to the experimental value of $\xi^2$, the red points are 50 Monte Carlo events generating new values $\xi^2_{mc}$. The three lines represent the critical values for a $\chi^2$ distribution for $p=0.05$ (dotted line), $p=0.01$ (dashed line), $p=0.005$ (solid line). The two figures correspond to $N_Y=200$ events (left panel), and to $N_Y=495$ events used to estimate $\Delta^2Y$, thus $\mathcal{L}$.} 
\label{fig:results}
\end{figure*}

However, the error on $\mathcal{L}$ makes this test incomplete.
We can not apply a Fisher-Snedecor test to compare two estimated variances, given that the reference value $1/(N_z\mathcal{L})$ is not $\chi^2$ distributed. We have then opted for a Monte Carlo approach, generating multiple values  $\mathcal{L}_{mc}$ by bootstrapping on the observed statistics, under the assumption the counts follow a Poisson law. Correspondingly, a new value $\xi^2_{mc}$ is evaluated and compared to critical values. The results are presented in Fig.~\ref{fig:results}. At lower values of $N_Z$ the violation of bound does not bear sufficient statistical significance, but this improves up to the $p=0.005$ confidence level.


\section{Conclusions}
 
We have shown that a direct assessment of YFG's criterion for quantum steering based on phase estimation demands overcoming some conceptual and practical challenges. These can be tackled by recurring to the multiparameter scenario as a strategy to gain robustness against imperfections. Nevertheless, it should be observed that the partitioning of information among the different parameters, including nuisances, increases the uncertainty on the target parameter $\theta$ as the test is stricter that in the single-parameter case. Consequently, hypothesis testing provides an effective framework for a rigorous assessment.

Thanks to the introduction of limits holding far from the asymptotic conditions of the Cram\'er-Rao bound, our approach opens up perspectives to explore the verification of quantum steering  based on the YFG paradigm in those contexts not allowing to correct large datasets, such as those involving multiple parties.

In particular, such an extended YFG approach can be relevant for distributed quantum sensing. Its appeal is in the fact that the same framework can be used for the estimation as well as for the certification of the channel. In particular, parties are explicitly allowed to place confidence in their apparatus, a standard assumption in quantum metrology. The extra requirements on the measurement devices to include the experimental certification can be worked out, and, thanks to multiparameter methods, imperfections can be largely taken into account. Our approach is a first step towards establishing the viability of this strategy.

\begin{backmatter}
\bmsection{Funding}
This work was supported by the PRIN project PRIN22-RISQUE-2022T25TR3 of the Italian Ministry of University.
G.B., F.B., and G.C. are supported by Rome Technopole Innovation Ecosystem (PNRR grant M4-C2-Inv). IG acknowledges the support from MUR Dipartimento di Eccellenza 2023-2027.

\bmsection{Acknowledgments}
We thank Adele Erculei and Nicolas Treps for stimulating discussion.

\bmsection{Disclosures}
The authors declare no conflicts of interest.

\bmsection{Data Availability Statement}
Data can be obtained by reasonable request to the corresponding author.


\end{backmatter}

\section*{Supplementary Material}
\subsection*{Fisher information}
In any parameter estimation problem~\cite{ParisReview}, in which a parameter $\theta$ must be learned from $N$ events extracted from the distribution $p(r|\theta)$,  the variance of the estimator is limited from below according to the Cram\'er-Rao bound
\begin{equation}
\label{eqs:fisingle}
    \text{Var}[\theta] f(\theta)\geq \frac{1}{N},  
\end{equation}
with the Fisher Information defined as 
\begin{equation}
    f(\theta)=\sum_r \frac{\left(\partial_\theta p(r\vert \theta)\right)^2}{p(r|\theta)}.
    \label{eqn:fisher_1p}
\end{equation}
If $p(r\vert \theta)$ describes the outcomes of measuring a quantum observable $R$ on a state $\rho$, then the information $f(\theta,\rho,R)$ is maximized by a quantity going under the name of  quantum Fisher information $f_Q(\theta,\rho)$: $f(\theta,\rho,R)\leq f_Q(\theta,\rho)$ for any $R$. It is also guaranteed that an optimal observable $\bar R$ exists achieving saturation. When $\theta$ is associated to the transformation $e^{-i\theta H}$ with generator $H$, then $f_Q(\theta,\rho)=4\Delta^2H$, the variance of $H$ evaluated on the state $\rho$.

These results are extended to the multiparameter scenario in which a vector of parameters $\vec \theta=\{\theta_1, \theta_2, ..., \theta_P \}$ must be evaluated. The covariance matrix $\text{Cov}[\vec \theta]$ of the parameters can be bound as
\begin{equation}
    \label{eqn:Fisher_information_matrix}
    \text{Cov}[\vec \theta]\geq \frac{1}{N}F(\vec \theta)^{-1} 
\end{equation}
by introducing the Fisher information matrix with elements
\begin{equation}
    F_{i,j}(\vec \theta)=\sum_r \frac{\left(\partial_{\theta_i} p(r\vert \vec \theta) \partial_{\theta_j} p(r\vert \vec \theta)\right)}{p(r|\vec \theta)}.
\end{equation}
The individual variances satisfy
\begin{equation}
\label{eqs:fimulti}
    \text{Var}[\theta_i]\geq \frac{1}{N}\left(F(\vec \theta)^{-1}\right)_{i,i}.
\end{equation}
The relations \eqref{eqs:fisingle} and \eqref{eqs:fimulti} are known to be strict only in the asymptotic condition of a large sample size $N$.  Outside this regime, one has to consider explicitly the prior knowledge  on the parameters by adopting a suitable distribution $\mathcal{P}(\vec \theta)$. If this is regular, then an inequality due to Van Trees holds: given the matrix with elements
\begin{equation}
\label{eqs:vt}
    V_{i,j}(\theta)= \int \mathcal{P}(\vec \theta) F_{i,j}(\vec \theta) d^P \theta+\frac{1}{N} \int  \frac{\left(\partial_{\theta_i} \mathcal{P}(\vec \theta) \partial_{\theta_j} \mathcal{P}(\vec \theta)\right)}{\mathcal{P}(\vec \theta)} d^P \theta
\end{equation}
then 
\begin{equation}
\label{eqs:vtbound}
    \text{Cov}[\vec \theta]\geq \frac{1}{N}V(\vec \theta)^{-1}.
\end{equation}
The matrix $V$ contains two terms, one associated to the average Fisher information, the second to the informational content of the prior distribution itself via its score matrix: as the number of copies increases, the latter becomes irrelevant and the expression \eqref{eqs:vtbound} reduces to the standard bound \eqref{eqs:fimulti}.

\subsection*{Extending Yadin, Fedel, and Gessner's treatment}

In any bipartite state shared by Alice and Bob,  the inference variance $\Delta^2 G_{\rm est}$ for Bob's observable $G$ can be bound from below as 
\begin{equation}
\label{eqs:deltaH}
    \Delta^2 G_{\rm est}\geq \Delta^2 G_{\rm min}=\min_K \sum_k p(k|K)\Delta^2G[\rho^B_{k|K}]
\end{equation}
where $\Delta^2G[\rho^B_{k|K}]$ is the ordinary quantum variance of $G$ evaluated in the state $\rho^B_{k|K}$. The  Fisher Information $f(\theta,\rho^{AB},K\otimes G')$ pertaining to a joint measurement of $K\otimes G'$  obeys the relation 
\begin{equation}
    f(\theta,\rho^{AB},K\otimes G') \leq f_{\rm max}(\theta)=\max_K \sum_k p(k|K)f(\theta,\rho^B_{k|K},G').
\end{equation}
The main result of YFG is that, for states admitting a local hidden state model
\begin{equation}
\label{eqs:yfg}
    f_{\rm max}(\theta) \leq 4\Delta^2 G_{\rm min},
\end{equation}
The Cram\'er-Rao bound then implies
\begin{equation}
    \text{Var}[\theta] \geq \frac{1}{N f_{\rm max} (\theta)} \geq \frac{1}{4N \Delta^2 G_{\rm min} },
\end{equation}
and, by using \eqref{eqs:deltaH},
\begin{equation}
\label{eqs:final}
    \text{Var}[\theta] \Delta^2 G_{\rm est} \geq \frac{1}{4N  }.
\end{equation}

In the case of a two-parameter estimation problem, with a target parameter $\theta$ and a nuisance parameter $v$, the Fisher information matrix satisfies
\begin{equation}
\label{eqs:comparison}
    \frac{1}{\left(F^{-1}\right)_{\theta,\theta}} = F_{\theta,\theta}- \frac{F_{\theta,v}^2}{F_{v,v}}\leq F_{\theta,\theta},
\end{equation}
given that $F_{\theta,v}^2$  and $F_{v,v}$ are positive quantities. We now recognise that $F_{\theta,\theta}=f(\theta)$, the Fisher information one would have available if the problem was an instance of single-parameter estimation. Adopting a similar chain of inequalities as before then leads to conclude \eqref{eqs:final} still holds in this two-parameter regime.
In the general case,  we  identify a target parameter $\theta_1$, and the vector of nuisance parameters $\vec v$ . The corresponding Fisher matrix is in block form
\begin{equation}
    F=\begin{pmatrix}
        F_{\theta,\theta}&\vec f_{\theta,\vec v}\\
        f_{\theta,\vec v}^\intercal& F_{\vec v, \vec v}
    \end{pmatrix},
\end{equation}
where the vector $\vec f_{\theta, \vec v}$ contains the correlation terms of $\theta$ with the nuisances, and $F_{\vec v,\vec v}$ is their Fisher information matrix. The expression of the inverse matrix $F^{-1}$ leads to
\begin{equation}
\label{eqs:generic}
    \frac{1}{\left(F^{-1}\right)_{\theta,\theta} }= F_{\theta,\theta}- \vec f_{\theta,\vec v}^\intercal \cdot F_{\vec v,\vec v}^{-1} \cdot \vec f_{\theta,\vec v}\leq F_{\theta, \theta},
\end{equation}
since $F_{\vec v,\vec v}$, being a valid Fisher information matrix, is non-negative. 
 
When introducing a prior distribution  $\mathcal{P}(\theta)$, we can still conclude, in perfect analogy with \eqref{eqs:yfg}, that the Van Trees inequality implies that 
\begin{equation}
    \label{eq:VT_bound}
        \text{Var}[\theta]\geq \frac{1}{N V_{\theta,\theta}},
\end{equation}
with
\begin{equation}
    V_{\theta,\theta} = \int \mathcal{P}(\theta,v)F_{\theta,\theta}(\theta,v)d\theta dv+\frac{1}{N}\int \frac{\left( \partial_\theta \mathcal{P}(\theta,v)\right)^2}{\mathcal{P}(\theta,v)} d\theta d v.
\end{equation}
In a local hidden state model, the Fisher information on $\theta$ can not exceed the optimum value $4\Delta^2G_{\rm min}$.  established by independent measurements. Therefore, it results
\begin{equation}
    V_{\theta,\theta}\leq 4\Delta^2 G_{\rm min}+\frac{1}{N} \int \frac{\left( \partial_\theta \mathcal{P}(\theta,v)\right)^2}{\mathcal{P}(\theta,v)} d\theta d v.
\end{equation}
A bound on the variance is then established
\begin{equation}
        N\text{Var}[\theta] \mathcal{L} \geq 1,
        \label{eq:bizzarri1}
\end{equation}
where $\mathcal{L}$ is defined in the main text.

\subsubsection*{Single-parameter scenario}
We provide here an effective model capturing the essential features of our experiment. The density matrix of a partially coherent singlet state is 
\begin{equation}
    \label{eqn:pc_singlet state}
    \rho_{AB}(v) = \frac{1}{2}\begin{pmatrix}
        0 & 0 & 0 & 0\\
        0 & 1 & -v & 0\\
        0 & -v & 1 & 0\\
        0 & 0 & 0 & 0
    \end{pmatrix}.
\end{equation}
This is equivalent to the state used in the experiment, up to local rotations, but has a simpler expression which improves the clarity of our treatment.  According to Alice's measurement outcomes of $X$, Bob's conditional state are, respectively
\begin{eqnarray}
    \label{eqn:conditional_states_DA}
    &\rho_D(\phi|v) = \rho^B_{D|X} = \frac{1}{4}\begin{pmatrix}
        1+v\sin(\phi) & -v\cos(\phi)\\
        -v\cos(\phi) & 1-v\sin(\phi)
    \end{pmatrix}\nonumber\\
    &\rho_A(\phi|v) = \rho^B_{A|X} =  \frac{1}{4}\begin{pmatrix}
        1-v\sin(\phi) & v\cos(\phi)\\
        v\cos(\phi) & 1+v\sin(\phi)
    \end{pmatrix}.
\end{eqnarray}

Therefore, conditional probabilities for Bob's measurement outcome $b$ given Alice's results $a$ can be retrieved by means of the usual Born's rule $p(b|a,K) = \frac{\Tr\left(\ketbra{b}{b}\rho_{a|K}^{B}\right)}{\Tr\left(\rho_{a|K}^{B}\right)}$. Such conditional probabilities enter the definition of the conditional Fisher information, {\it viz.}
 \begin{equation}
     \label{eqn:CFI}
     f_{B|A}(\phi|v) = \sum_{a}p(a|K)f_{a|K}(\phi|v),
 \end{equation} 
where $f_{a|K}(\theta|v)$ represents  the Fisher information \eqref{eqn:fisher_1p}  available to Bob when Alice's measurement result is $a$. The explicit calculation using Bob's conditional states \eqref{eqn:conditional_states_DA} gives
 \begin{equation}
    \label{eqn:CFI:_singlet}
    f_{B|A}(\phi|v) = \frac{v^2\cos^2(\phi)}{1-v^2\sin^2(\phi)}.
 \end{equation}
 having its maximum value at $\phi = m\pi,\, m \in \mathbb{Z}$ :$f_{max}= f_{B|A}(\phi = m\pi|v) = v^2$.
\begin{figure}[h]
\includegraphics[width=\columnwidth]{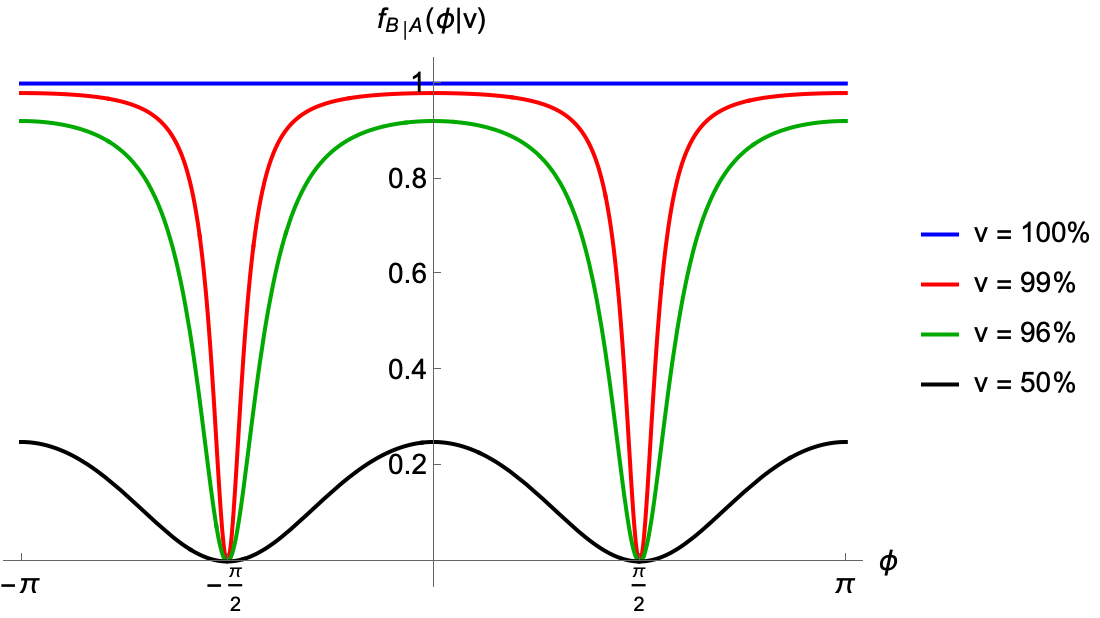}
\caption{Behavior of the conditional Fisher information in the $[-\pi,\pi]$ interval, for different values of the nuisance parameter $v$ .
}
\label{fig:fisher_1p}
\end{figure}\noindent\\
 Similarly, Bob's conditional states given Alice's measurement of Y are
 \begin{equation}
     \label{eqn:conditional_states_LR}
     \begin{cases}
         \rho_L(v) = \rho_{L|Y}^{B} = \frac{1}{4}\begin{pmatrix}
             1 & iv\\
             -iv & 1
         \end{pmatrix}\\\\
         \rho_R(v) = \rho_{R|Y}^{B} = \frac{1}{4}\begin{pmatrix}
             1 & -iv\\
             iv & 1
         \end{pmatrix}.
     \end{cases}
 \end{equation}
 These allow to gather information about the generator $G = Y/2$, whose conditional variance is
 \begin{equation}
     \label{eqn:conditional_variance}
     \Delta^2_{B|A}Y = \sum_{a}^{R,L}{p(a|Y)\Delta^2_{a}Y},
 \end{equation}
 which YFG ~\cite{YBG} use to establish a limit to the saturation of the CRB:
 \begin{equation}
     \label{eqn:YFG_claim}
     \max_{\theta}f_{B|A}(\theta) \leq \Delta^2 H_{\rm est} \leq 4\Delta_{B|A}^2H,
 \end{equation}
 for all states obeying an LHS model.
 
 Given the state \eqref{eqn:pc_singlet state}, it is easy to verify that
 \begin{equation}
     \Delta_{B|A}^2Y = 1-v^2,
     \label{eqn:conditional_variance_singlet}
 \end{equation}
since $\Delta_a^2Y = \ev{Y^2}_a - \ev{Y}_a^2$, with $\ev{Y^2}_a = \ev{Y}_a = 1$, and $\ev{Y}_a = \frac{\Tr\left(Y\rho_{a|Y}^{B}\right)}{\Tr\left(\rho_{a|Y}^{B}\right)} = v$. When $v$ is a known parameter, a single-parameter treatment is adequate to look for a violation of  YFG's condition \eqref{eqn:YFG_claim} . By means of   \eqref{eqn:CFI:_singlet} and \eqref{eqn:conditional_variance_singlet},  we calculate that this is equivalent to 
\begin{equation}
     \label{eqn:YFG_ineq_singlet}
     v^2 > 1-v^2,
 \end{equation}
whose solution yields a visibility range $\frac{1}{\sqrt{2}}<v\leq1$ in which \eqref{eqn:YFG_claim} is violated.
 \begin{figure}[h]
     \centering
     \includegraphics[width=\columnwidth]{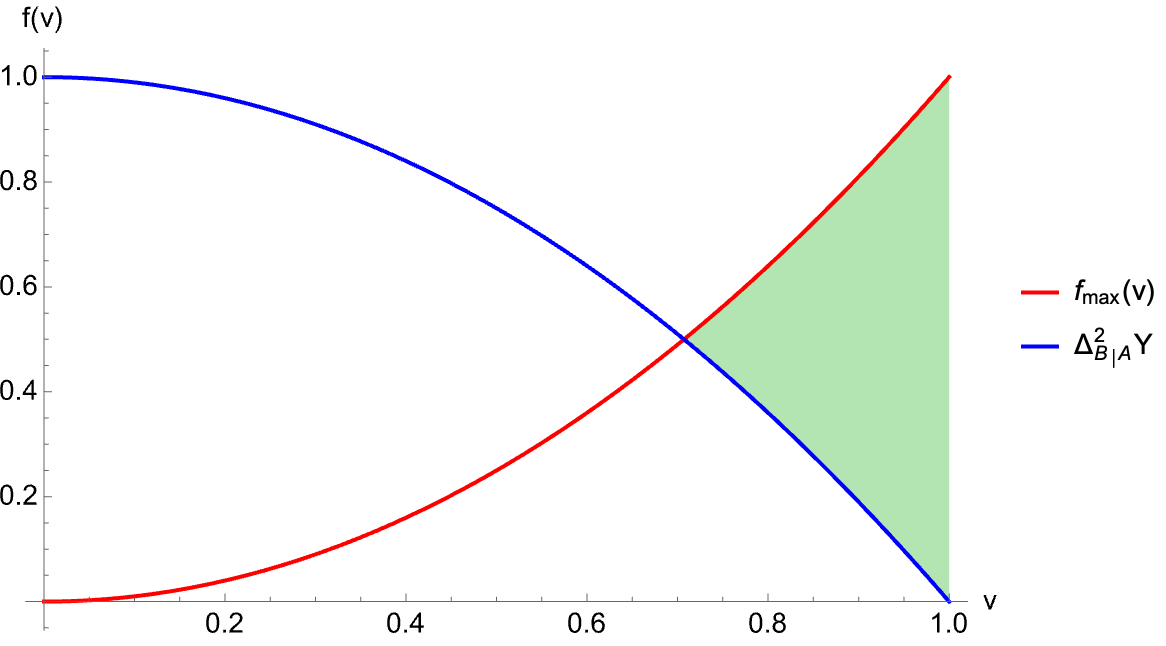}
     \caption{Plot of the maximal Fisher information $f_\text{max}$ and $4\Delta_{B|A}^2H = \Delta_{B|A}^2Y$ as a function of the nuisance parameter $v$. The green shaded area represent the visibility range in which YFG's inequality \eqref{eqn:YFG_claim} is violated (or, equivalently, inequality \eqref{eqn:YFG_ineq_singlet} is abide.}
     \label{fig:YFG_violation_1p}
 \end{figure}\\\noindent
If we want to include the role of Bob's {\it a priori} information, we should use his prior distribution $\mathcal{P}(\theta)$, appearing in the definition of the Van Trees' bound. In the single parameter case, Van Trees' inequality is reduced to a scalar formulation as in \eqref{eqs:generic}. Similarly, by exploiting the YFG inequality \eqref{eqs:yfg},  the following inequality is retrieved
 \begin{equation}
     \label{eqn: VT-YFG_1p}
     \frac{V(N|v)-V_{\text{YFG}}(N|v)}{NV(N|v)V_{\text{YFG}}(N|v)}\leq0,
 \end{equation}
with
\begin{equation}
    \label{eqn: V_YFG_1p}
    V_{\text{YFG}}(N|v) = \mathcal{P}(\phi)\Delta_{B|A}^2Y+\frac{1}{N}\int \frac{\left( \partial_\phi \mathcal{P}(\phi,v)\right)^2}{\mathcal{P}(\phi,v)} \dd\phi.
\end{equation}

\begin{figure}[!h]
    \centering
    \includegraphics[width = \columnwidth]{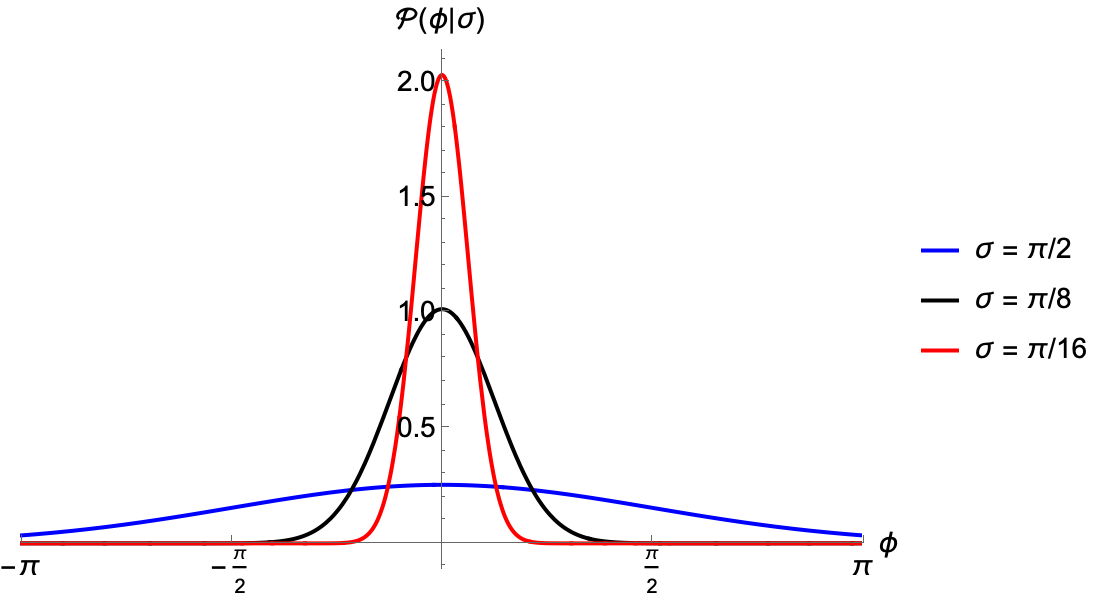}
    \caption{Plot of the phase prior distribution \eqref{eqn:prior_1p} in the $[-\pi,\pi]$ interval, centered at $\theta = 0$, for different values of the standard deviation $\sigma$.}
    \label{fig:phase_prior_1p}
\end{figure}

\begin{figure}[h]
    \centering
    \includegraphics[width = \columnwidth]{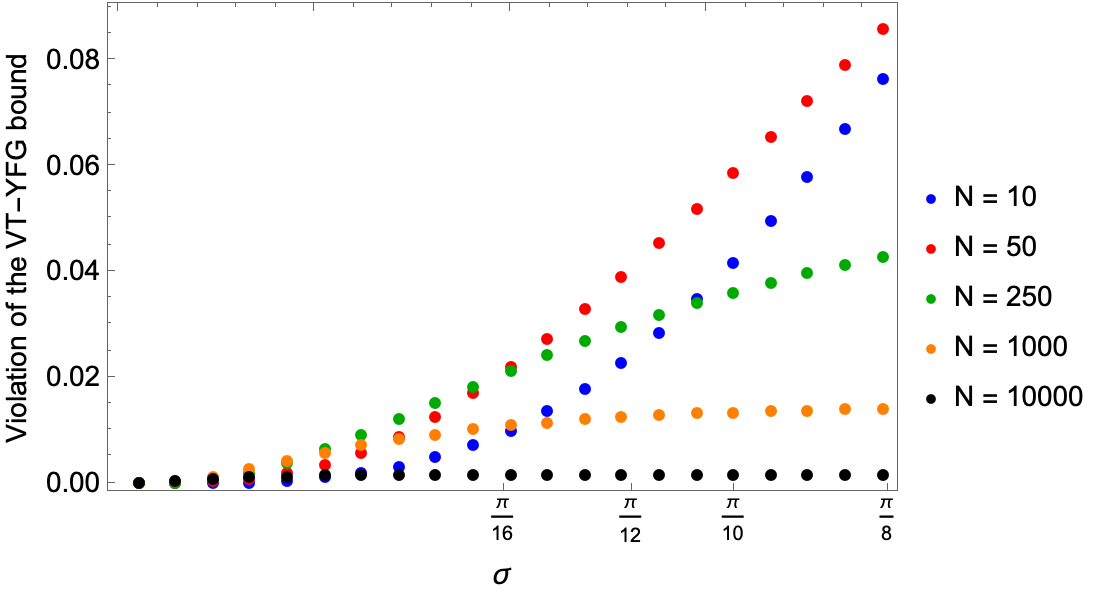}
    \caption{Violation of the VT-YFG inequality \eqref{eqn: VT-YFG_1p} at visibility $v = 97\%$, as function of the prior's standard deviation $\sigma$, for different values of N.}
    \label{fig:violation_1p_s}
\end{figure}

We used a Gaussian phase prior distribution
\begin{equation}
    \label{eqn:prior_1p}
    \mathcal{P}(\phi|\sigma) = \frac{e^{-\frac{\theta^2}{2\sigma^2}}}{\sqrt{2\pi\sigma^2}},
\end{equation}
as shown in figure \ref{fig:phase_prior_1p}, centered at the maximal informative phase, with $\sigma \leq \pi/8$ in order to disambiguate among different phases having the same conditional Fisher information: we assume Bob already knows at which fringe his apparatus is operating.
\\\noindent
The violation of the VT-YFG inequality \eqref{eqn: VT-YFG_1p} depends on both the number of resources N used to infer the value of the phase $\theta$ and the prior's standard deviation $\sigma$.
As shown in Fig.~\ref{fig:violation_1p_s}, the more resources are employed, the less sensitive the violation of \eqref{eqn: VT-YFG_1p} is to variations of the prior's standard deviation.
\\\noindent
At any given $\sigma$, the absolute violation of the VT-YFG inequality does not necessarily grow with the number $N$ of resource, since the limited covariances are reduced proportionally: an example is reported in Fig.~\ref{fig:violation_1p_s}. A larger absolute violation can actually be less statistical significant, when it is ascertained with a reduced number of degrees of freedom for the associated distribution.



\subsubsection*{Multiparameter scenario}
\begin{figure}[h!]
    \centering
    \subfloat[][First diagonal element of the inverse \\conditional Fisher information matrix
    \label{fig:Finv_11}]{\includegraphics[width = .5\textwidth]{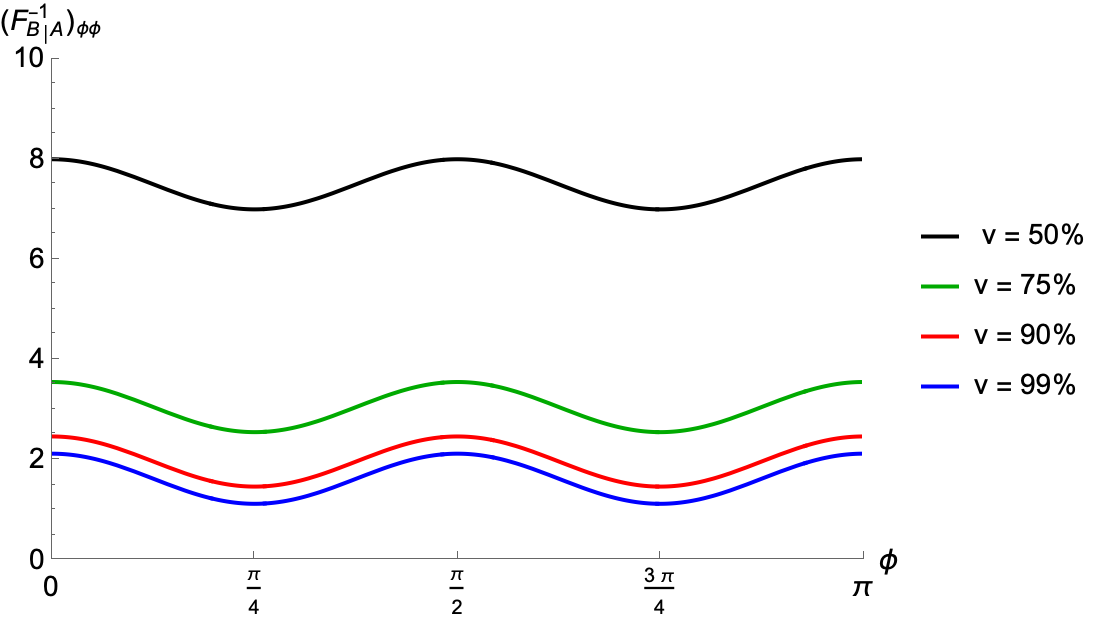}}\hfill
    \subfloat[][Second diagonal element of the inverse \\conditional Fisher information matrix\label{fig:Finv_22}]{\includegraphics[width = .5\textwidth]{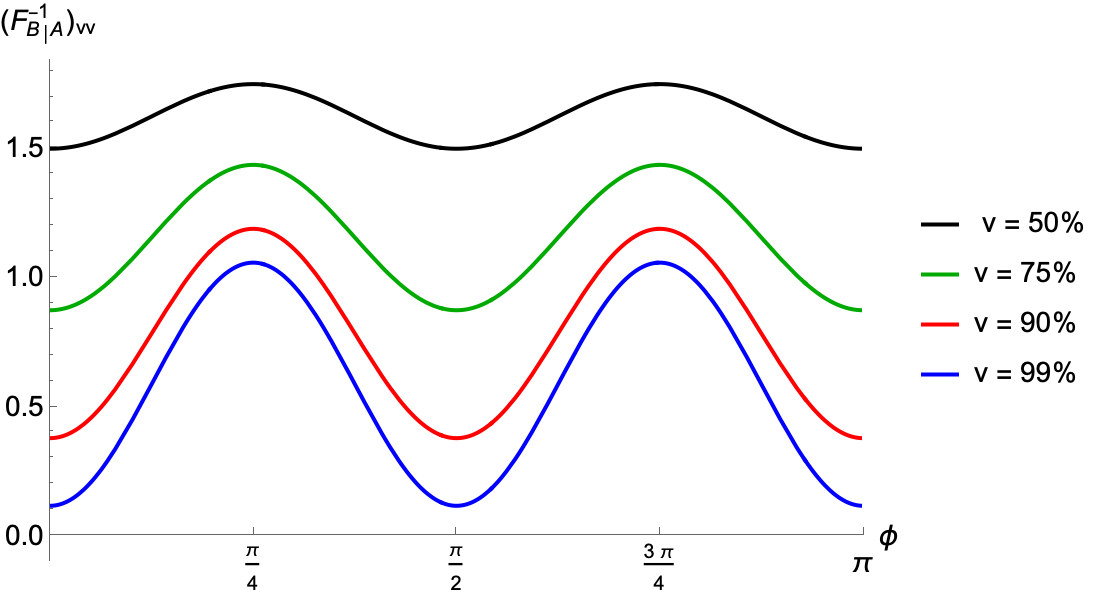}}\\
    \subfloat[\centering][Correlation coefficient $R(\phi) = \frac{\left(F^{-1}_{B|A}\right)_{\phi v}}{\sqrt{\left(F^{-1}_{B|A}\right)_{\phi\phi}\left(F^{-1}_{B|A}\right)_{vv}}}$\label{fig:Corr_coeff}]{\includegraphics[width = .5\textwidth]{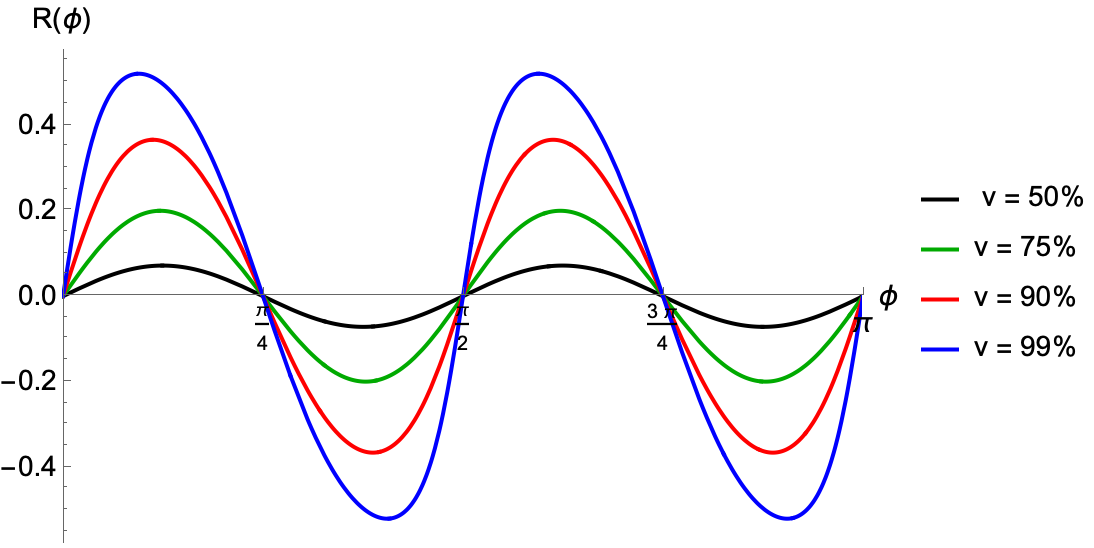}}
    \caption{Plots of the matrix element of the inverse conditional Fisher information matrix as function of the phase $\phi$, at given values of the nuisance parameter $v$. Figs. ~\ref{fig:Finv_11} and ~\ref{fig:Finv_22} represent the lower bound for the estimation error on, respectively, the phase shift $\phi$ and the coherence term $v$; whereas ~\ref{fig:Corr_coeff} describes how correlations between $\phi$ and $v$ vary with respect to the uncertainties on the aforementioned parameters.}
    \label{fig:Finv_plots}
\end{figure}

When moving to the multiparameter scenario, the relevant metrological bound are expressed in terms of the inverse Fisher information matrix \eqref{eqn:Fisher_information_matrix}. The conditional Fisher information matrix pertaining to our model is
\begin{equation}
    F_{B|A} = \begin{pmatrix}
       \frac{v^2}{2}\left(\frac{\cos^2(\phi)}{1-v^2\sin^2(\phi)}+\frac{\sin^2(\phi)}{1-v^2\cos^2(\phi)}\right) & -\frac{v^3\sin(4\phi)}{8(1-v^2)+2v^4\sin^2(2\phi)}\\ 
       -\frac{v^3\sin(4\phi)}{8(1-v^2)+2v^4\sin^2(2\phi)} & \frac{\sin^2(\phi)}{2(1-v^2\sin^2(\phi))}+\frac{\cos^2(\phi)}{2(1-v^2\cos^2(\phi))}
    \end{pmatrix}
    \label{eqn:conditional_FIM}
\end{equation}
The matrix elements of its inverse - the quantities actually relevant to the metrological bounds - are shown in Fig. ~\ref{fig:Finv_plots}.
\\\noindent
The optimal working conditions are found around $\phi=\pi/4$ (and its multiples): the bound on its precision is the lowest, thus making it harder for a malicious Alice to replicate the performance of the state, and the correlation with the visibility estimation is vanishing.  

\begin{figure}[h]
    \centering
    \includegraphics[width = \columnwidth]{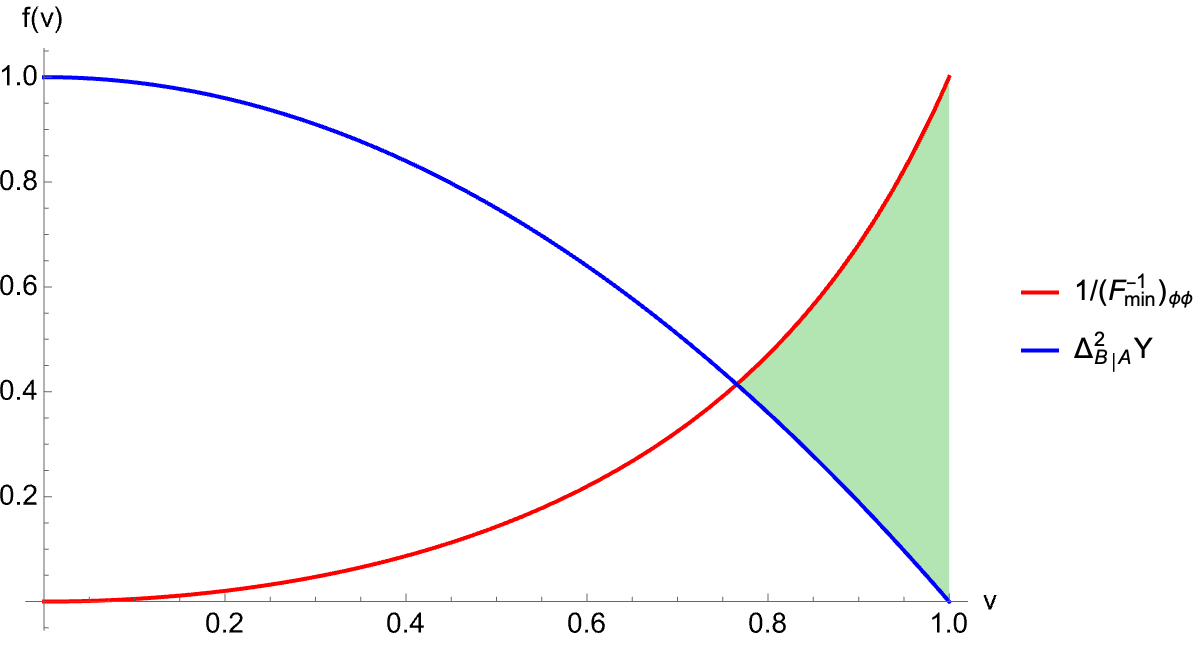}
    \caption{Plot of the maximal effective Fisher information $1/(F_{\text{min}}^{-1})_{\phi\phi}$ and $\Delta_{B|A}^2Y$ as a function of the nuisance parameter $v$. The green shaded area represent the visibility range $(\sqrt{2-\sqrt{2}},1]$ in which YFG's inequality \eqref{eqn:multip_YFG} is violated.}
    \label{fig:violation_YFG_2p}
\end{figure}

\begin{figure}[!h]
    \centering
    \includegraphics[width =\columnwidth]{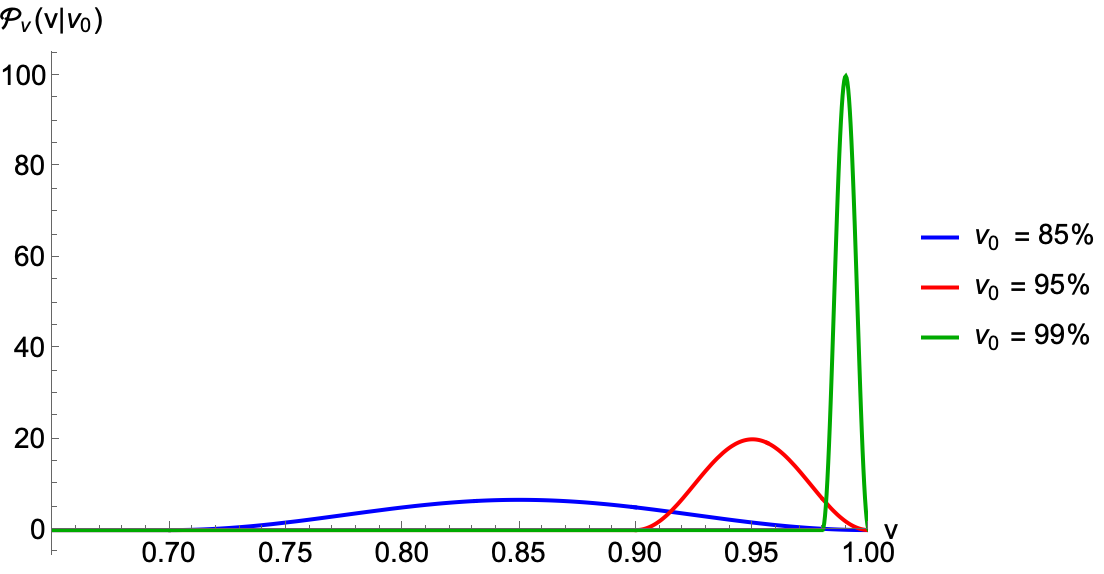}
    \caption{Plot of the prior distribution $\mathcal{P}_v(v|v_0)$ on the nuisance parameter, for different values of $v_0$ in the violation range $(\sqrt{2-\sqrt{2}},1]$.}
    \label{fig:visibilty_prior}
\end{figure}

Like for the single parameter scenario, YFG's approach bounds the conditional variance of the generator with the variance of the conjugate phase shift, now constrained by the multiparameter Cram\'{e}r-Rao bound \eqref{eqs:fimulti}. By combining it with the YFG inequality \eqref{eqs:yfg} , the following lower bound is retrieved
\begin{equation}
    \Delta^2_{B|A}Y\geq\frac{1}{N(F^{-1}_{\text{min}})_{\phi\phi}}
    \label{eqn:multip_YFG}
\end{equation}
where $(F^{-1}_{\text{min}})_{\phi\phi}=F^{-1}_{\phi\phi}(\phi = \pi/4,v)= \frac{2-v^2}{v^2}$, while the conditional variance of the generator is the same as in the single parameter scenario. The graphical representation of the violation of the multiparameter YFG inequality is shown in Fig.\ref{fig:violation_YFG_2p}
\\\noindent
When dealing with a limited-data regime in the multiparameter framework, the prior distribution $\mathcal{P}(\phi,v)$ encoding Bob's initial knowledge must account for the nuisance parameter as well.  At $\phi = \pi/4$, there are no correlations between the estimation of the nuisance parameter $v$ and of the phase $\phi$: thus, it is convenient to take a prior distribution that factorizes into individual distributions, {\it i.e.} $\mathcal{P}(\phi,v) = \mathcal{P}_{\phi}(\phi|\sigma)\mathcal{P}_v(v|v_0)$.  The phase prior distribution can still be taken in the form of \eqref{eqn:prior_1p},now centered at $\phi = \pi/4$ and $\sigma \leq \pi/16$ because of the halved periodicity of the Fisher information matrix elements. The prior distribution on the nuisance parameter $v$ must be be chosen as a probability function with support in the interval $v \in [0,1]$, obeying the regularity constraints imposed by the Van Trees' lower bound. A suitable choice is a raised cosine distribution of the form
\begin{equation}
    \label{eqn:vis_prior}
    \mathcal{P}_v(v|v_0) = \frac{\Theta(v-(2v_0-1))}{2(1-v_0)}\left(1+\cos\left(\pi\frac{v-v_0}{1-v_0}\right)\right),
\end{equation}
where $v_0$ corresponds to the expectation value of the nuisance parameter $v$, and $\Theta(x)$ is Heaviside's step function . Such a probability distribution is non-zero in the interval $(2v_0-1,1)$ , and is represented in Fig.~\ref{fig:visibilty_prior}.  It is desirable that  $v_0$ should lie fall the violation range $(\sqrt{2-\sqrt{2}},1]$.
The variance on the phase $\phi$ is lower bounded by the multiparameter Van Trees' inequality \eqref{eqs:vtbound}, namely
\begin{equation}
    \label{eqn:multip_vtbound_phase}
    \text{Var}\left[\phi\right] \geq \frac{V^{-1}_{\phi\phi}}{N},
\end{equation}
where $V^{-1}_{\phi\phi}$ denotes the first element of the inverse Van Trees' matrix, in perfect analogy with the standard multiparameter Cram\'er-Rao bound. Inserting this condition in \eqref{eq:bizzarri1}, the following bound is retrieved
\begin{equation}
    \label{eqn:multip_VTYFG_ineq}
    V^{-1}_{\phi\phi}\mathcal{L}\geq 1,
\end{equation}
a violation of which states the unsuitability of a LHS model, assessing then the presence of quantum steering.\\
Analogously to the single parameter case, the validity of equation \eqref{eqn:multip_VTYFG_ineq} is conditioned on  the free parameters of the prior distribution, namely $\sigma$ and $v_0$.\\
\begin{figure}[!h]
    \centering
    \includegraphics[width = \textwidth]{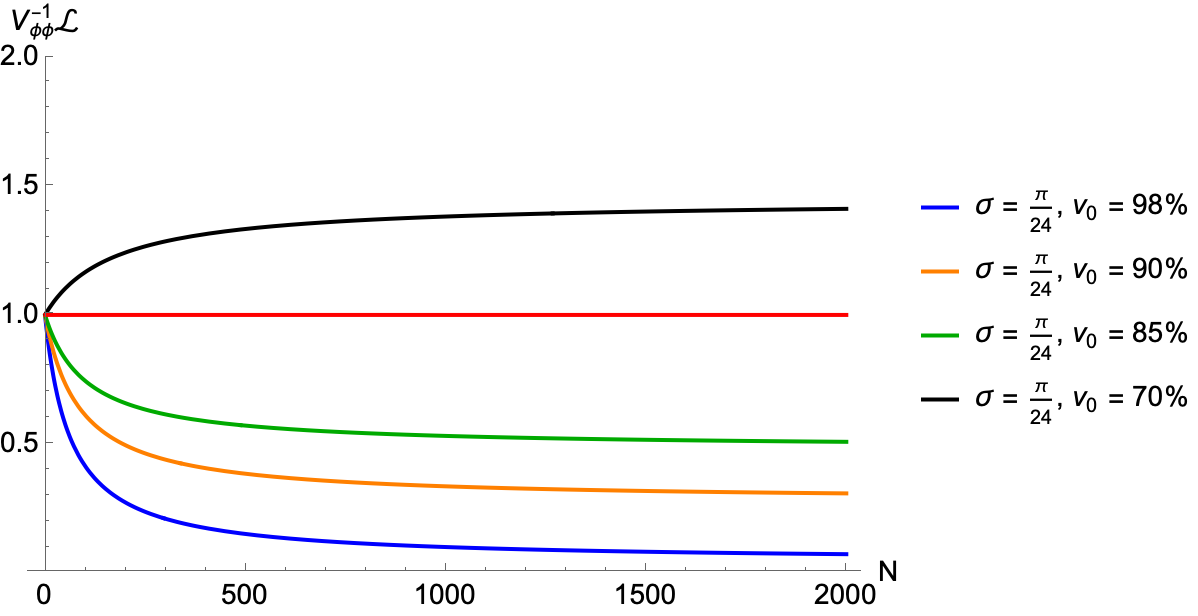}
    \caption{Violation of \eqref{eqn:multip_VTYFG_ineq} as a function of the number $N$ of resourced, at fixed $\sigma$, for different values of the average visibility $v_0$.}
    \label{fig:VTYFG_2p_violation_v0}
\end{figure}
Figure ~\ref{fig:VTYFG_2p_violation_v0} highlights that, whenever $v_0$ lies within the visibility's violation range $(\sqrt{2-\sqrt{2}},1]$, the VT-YFG inequality \eqref{eqn:multip_VTYFG_ineq} is violated for any value of $N$, with the asymptotic value getting closer to $1$ as the average visibility decreases.


 \begin{figure}[!h]
    \centering
    \includegraphics[width = \textwidth]{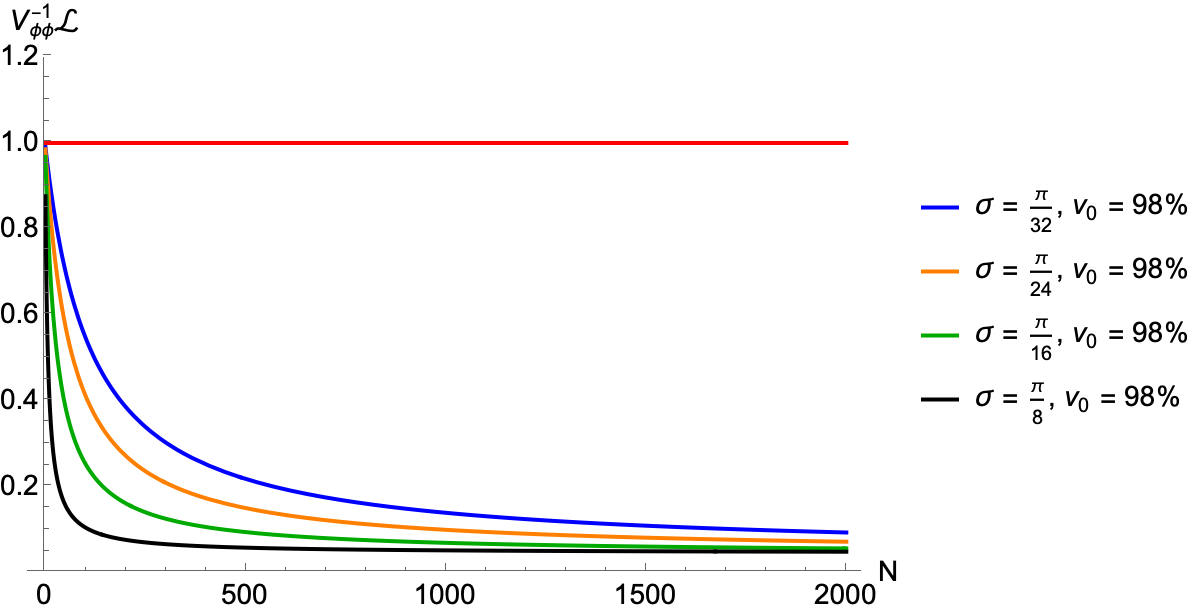}
    \caption{Violation of \eqref{eqn:multip_VTYFG_ineq} as a function of the number $N$ of resourced, at fixed $v_0$, for different values of the phase prior's standard deviation $\sigma$.}
    \label{fig:VTYFG_2p_violation_s}
\end{figure}

\subsection*{Data analysis}
As discussed in the main text, depending on Alice's measurement setting, Bob can decide to either  perform the joint estimation of phase and visibility, or to assess the variance $\Delta^2_{B|A}Y$, so as to establish a bound on the available Fisher information. More specifically, in the first scenario, the sum of all the occurrences associated to any combination of Alice's and Bob's measurements must be equal to the number $N_Z$ of physical resources addressed to such task
\begin{equation}
    \label{eq:NZ_partition}
    \sum_{b}\sum_{a}n_{ba} = N_Z \qquad b = D,A,H,V;\; a = H,V,
\end{equation}
corresponding to the number of Alice's interrogations of Z. Similarly, in the second instance, the occurrences $N_Y$ of Alice measuring $Y$ are arranged among all possible combinations of results of $Y$ by both Alice and Bob
\begin{equation}
    \label{eq:NY_partition}
    \sum_{b,a}{m_{ba}} = N_Y\qquad b = L,R;\; a = L,R.
\end{equation}
Here we detail how we have employed these events for our data analysis.

\subsubsection*{Bayesian estimation of the parameters}
Given $N_Z$ independent interrogations of  $Z$ by Alice, the probability of observing a given result $b$ on Bob's side, given the result $a$ by Alice, will be proportional to $p(b|a)^{n_{ba}}$, with $p(b|a)$ being the corresponding probability. The collection of all of Bob's results allows then to define a \textit{conditional likelihood function}
\begin{equation}
    \label{eq:conditional_likelihood}
    H_a = \prod_{b}p(b|a)^{n_{ba}}
\end{equation}
for any of Alice's result $H$ or $V$. These probabilities can be obtained from expressions similar to \eqref{eqn:conditional_states_DA}, adapted to our expected state, and are all in the form $\left(1+v\cos(\phi+\Gamma_{ab})\right)/4$, with $\Gamma_{ab}$ determined by the specific setting. This provides a description in terms of a model with decoherence.

The Bayesian approach is well suited to include the prior information in a straightforward way. In this frame, the estimator of the phase shift $\phi$ and  its variance $\text{Var}[\phi]$ for a given Alice's result $a$ are
\begin{equation}
    \label{eq:fixed_value_estimators}
    \begin{cases}
        \bar{\phi}_a =\frac{\int\phi\mathcal{P}(\phi,v)H_a(\phi_v)\dd\phi\dd v}{\int\mathcal{P}(\phi,v)H_a(\phi_v)\dd\phi\dd v}\\\\
        \Delta^2_a\phi = \frac{\int(\phi-\bar{\phi}_a)^2\mathcal{P}(\phi,v)H_a(\phi_v)\dd\phi\dd v}{\int\mathcal{P}(\phi,v)H_a(\phi_v)\dd\phi\dd v}
    \end{cases}
    \quad a = H, V.
\end{equation}
Similar equations hold for the Bayesian estimation of the nuisance parameter $v$ and its variance.

In the genuine spirit of YFG's method \cite{YBG}, the \textit{conditional Bayes estimators} for the phase $\phi$ and its variance are given by the linear combination of the estimators at a given Alice's measurement, weighted by the relative occurrence of that result (squared relative occurrence for the conditional variance)
    \begin{equation}
    \label{eq:Conditional_estimators}
    \begin{cases}
        \bar{\phi}_{\text{cond}} = \frac{\sum_b\sum_a n_{ba}\bar{\phi}_a}{\sum_b\sum_a n_{ba}}\\
        \\
        \Delta^2\phi_{\text{cond}} = \frac{\sum_b\sum_a n^2_{ba}\Delta^2_a\phi}{\left(\sum_b\sum_a n_{ba}\right)^2}
    \end{cases}
    \qquad a = H,V,
\end{equation}
where $n_a = \frac{\sum_b n_{ba}}{\sum_b\sum_a n_{ba}}$ is the relative occurrence for the result $a = H,V$.\\

\subsubsection*{Reconstruction of the generator}

Given the occurrences $m_{ba}$ of all the possible combinations of measurement outcomes of $Y$ by both Alice and Bob, obeying the condition \eqref{eq:NY_partition}; the measured value $\bar{Y}_a$ of Bob's observable $Y$ at a given Alice's outcome is retrieved directly as the average value of $Y$
\begin{equation}
    \label{eq:Yest_fixed result}
    \bar{Y}_a = \frac{m_{La}-m_{Ra}}{m_{La}+m_{Ra}}\qquad a = L,R;
\end{equation}
The result $1$ is attributed to the counterclockwise circular polarization $L$, and the result $-1$ to the clockwise circular polarization $R$. Hence, the conditional estimation of $Y$ is retrieved as
\begin{equation}
    \label{eq:Ycond_estimator}
    Y_{\text{cond}} = \frac{m_L}{N_Y}Y_L + \frac{m_R}{N_Y}Y_R,
\end{equation}
where $m_a = \sum_{b = L,R} m_{ba}\quad a = L,R$.\\

The conditional variance on the generator is then established as
\begin{equation}
    \label{eq:deltaYcond_est}
    \Delta^2 Y_{\text{cond}} = 1 - Y^2_{cond},
\end{equation}
retrieved, by recalling that $\sigma_j^2 = 1$ for all Pauli matrices.\\

\subsubsection*{Figures of merit}
The conditional variance of the phase estimator is lower bounded by the corresponding element of the inverse Van Trees' matrix \eqref{eq:VT_bound}: for each of Alice's measurement outcome $a = H,V$, a matrix can be constructed 
\begin{equation}
    \label{eq:VT_Alice_fixed}
    V^{(a)}_{\mu\nu} = \int \mathcal{P}(\phi,v)F_{\mu\nu}(\phi,v)\dd\phi \dd v +\frac{1}{n_a}\int \frac{\partial_\mu \mathcal{P}(\phi,v)\partial_\nu\mathcal{P}(\phi,v)}{\mathcal{P}(\phi,v)} \dd\phi \dd v
\end{equation}
with $a = H,V\; \mu,\nu = \phi,v$,  and these use a restricted number of resources $n_a = \sum_{b}n_{ba}$. The total Van Trees matrix is then retrieved by a  linear combination of such two terms weighted with the corresponding resources:
\begin{equation}
    \label{eq:conditional_VT_matrix}
    V_{\mu\nu} = n_H V^{(H)}_{\mu\nu}+n_VV^{(V)}_{\mu\nu}\qquad \mu,\nu = \phi,v.
\end{equation}
This is dictated by the additive charactred of the Fisher information.

The YFG criterion \cite{YBG} is applied by comparing the conditional variance of the phase estimator \eqref{eq:Conditional_estimators} with the figure of merit $\mathcal{L}$, defined in the main text, accounting for the conditional variance of the generator $Y$.\\Such a bound is experimentally calculated by using the estimator \eqref{eq:deltaYcond_est} for $4\Delta^2G_{\text{cond}}$, leading to the following expression:
\begin{equation}
    \label{eq:L_experimental}
    \mathcal{L} =  \Delta^2 Y_{\text{cond}} + \frac{1}{N_Z}\int \frac{\left( \partial_\phi \mathcal{P}(\phi,v)\right)^2}{\mathcal{P}(\theta,v)} \dd\theta \dd v,
\end{equation}
where $\Delta^2 Y_{\text{cond}}$ is reconstructed by using $N_Y$ physical resources, whereas the score is divided by $N_Z$, i.e. the physical resources spent to build the phase estimator.

\bibliography{nonasymYFG.bib}

\end{document}